\documentclass[12pt, a4paper, oneside]{article}
\usepackage[utf8]{inputenc}
\usepackage[english]{babel}

\usepackage{geometry}
\usepackage{pdflscape}
\usepackage{graphicx}
\usepackage{amsmath}
\usepackage{amssymb}
\usepackage{amsthm}
\usepackage{url}
\usepackage{color}
\usepackage{float}
\usepackage{natbib}
\usepackage{caption}
\usepackage{subcaption}
\usepackage{multirow}
\usepackage{bbm}
\usepackage[table,x11names]{xcolor}
\usepackage{hyperref}

\newtheorem{algo}{Algorithm}[section]
\newtheorem{defi}{Definition}[section]

\newtheorem{rem}{Remark}[section]
\newtheorem{example}{Example}[section]
\newtheorem{coro}{Corollary}[section]

\newtheorem{propo}{Proposition}[section]
\newtheorem*{theorem*}{Theorem}

\newcommand{\N}{\mathbb N}

\newcommand{\R}{\mathbb R}
\newcommand{\Z}{\mathbb Z}

\title{Volterra bootstrap: Resampling higher-order statistics for strictly stationary univariate time series}
\date{\today}
\author{Natalia Sirotko-Sibirskaya\footnote{University of Bremen, Institute for Statistics, Germany}, Matthias O. Franz\footnote{HTWG Konstanz, Germany}, and Thorsten Dickhaus\footnote{Corresponding Author. University of Bremen, Institute for Statistics, Bibliothekstr. 1, 28359 Bremen, Germany. E-mail: dickhaus@uni-bremen.de}}
\begin{document}

\maketitle

\begin{abstract}
We are concerned with nonparametric hypothesis testing of time series functionals. It is known that the popular autoregressive sieve bootstrap is, in general, not valid for statistics whose 
(asymptotic) distribution depends on moments of order higher than two, irrespective of whether the data come from a linear time series or a nonlinear one. Inspired by nonlinear system theory we circumvent this non-validity by introducing a higher-order bootstrap scheme based on the Volterra series representation of the process. In order to estimate coefficients of such a representation efficiently, we rely on the alternative formulation of Volterra operators in reproducing kernel Hilbert space. We perform polynomial kernel regression which scales linearly with the input dimensionality and is independent of the degree of nonlinearity. We illustrate the applicability of the suggested Volterra-representation-based bootstrap procedure in a simulation study where we consider strictly stationary linear and nonlinear processes. 
\end{abstract}

MSC 2020 classification numbers: Primary 
          62M15,  
					62F40;  
          secondary
          62M10.  
          
          JEL Classification: C22, C12.
          
          Key words:             autocorrelation,
					                       autoregressive processes, 
					                       higher-order cumulants, 
					                       hypothesis testing,
                                 nonlinearity,
																 reproducing kernel Hilbert space.

\section{Introduction}\label{intro}
\sloppy
Over the recent years the bootstrap procedure initially introduced by \cite{efron1979annals,efron1979computers} for stochastically independent and identically distributed (iid) observables has been extended to cope with dependent data, see the overviews by \cite{hardle2003bootstrap}, \cite{kreiss2011bootstrap} and \cite{kreiss2012bootstrap} as well as the monographs by \cite{politis1999subsampling} and \cite{lahiri2003bootstrap},  among others. Most of the suggested methods deal with linear processes and often the sample mean is the only statistic of interest. However, real data often exhibit nonlinear patterns and statistics of higher order such as autocovariances, autocorrelations and spectral densities are of considerable interest. This motivates us to introduce a bootstrap procedure which takes into account nonlinear features of the time series reflected in its higher-order moments and to consider scenarios where it is especially beneficial to take such nonlinear features into consideration.

Alongside with linear strictly stationary time series we consider nonlinear strictly stationary time series $(X_t)_t$ as described by \cite{wu2005nonlinear,wu2011asymptotic}. They are of the form 
\begin{align}\label{wu}
X_t = H(\hdots, \varepsilon_{t-1},\varepsilon_t),
\end{align}
where $\{\varepsilon_t, \ t \in \mathbbm{N} \}$ are iid random variables and $H$ is a measurable function such that $X_t$ is well-defined. As \cite{wu2005nonlinear} argues the representation in \eqref{wu} can be viewed as a nonlinear analogue of the Wold representation. However, whereas for the Wold decomposition to hold one needs weakly stationary time series, asymptotic theory established in \cite{wu2005nonlinear} under the representation as in \eqref{wu} requires the time series to be strictly stationary. 

Throughout this work, we consider the following representation of nonlinear time series:
\begin{align}\label{volterra}
X_t = H(\hdots, \varepsilon_{t-1},\varepsilon_t) = \sum_{p=0}^{\infty} \sum_{u_1, \hdots, u_p = 0}^{\infty} h^{(p)} \varepsilon_{t- u_1} \hdots \varepsilon_{t - u_p},
\end{align}
or, equivalently,  
\begin{eqnarray*}
X_t  &= &h^{(0)} + \sum_{u=0}^{\infty} h^{(1)} \varepsilon_{t-u} + \sum_{u=0}^{\infty} \sum_{v=0}^{\infty} h^{(2)} \varepsilon_{t-u} \varepsilon_{t-v} + \hdots\\ 
& &+ \sum_{u=0}^{\infty} \sum_{v=0}^{\infty} \hdots \sum_{w=0}^{\infty} h^{(p)} \varepsilon_{t-u} \varepsilon_{t-v} \hdots \varepsilon_{t-w} + \hdots,
\end{eqnarray*}
where $h^{(p)}$ is a Volterra operator of order $p \geq 0$ and $(\varepsilon_t)_t$ denotes a sequence of real-valued random variables. The representation in \eqref{volterra} is called (discrete time) Volterra series expansion, due to the Italian mathematician Vito Volterra who suggested a continuous-time analogue of this functional form in the 1880s. The Volterra representation can be thought of as a Taylor series type expansion, but unlike Taylor series Volterra series capture so-called memory effects of time series reflected in the lags of the $\varepsilon_{t}$'s. 

Representation as in Equations \eqref{wu} and \eqref{volterra} were studied by \cite{wiener1958nonlinear}, whose work plays an important role in the nonlinear system theory, see, e.\ g., \cite{schetzen2006volterra}, \cite{mathews2000polynomial}, and \cite{rugh1981}, among others. \cite{wiener1958nonlinear} conjectured that if the process is stationary and ergodic, then there exists a function $H$ and iid 
random variables $(\varepsilon_{t})_t$ such that \eqref{wu} holds. These conditions were, however, shown to be insufficient by \cite{rosenblatt2009comment}. On the other hand, it has been proven by \cite{nisio1960polynomial} that every strictly stationary time series has a two-sided polynomial representation in terms of Gaussian iid random variables. However, according to our knowledge the sufficient conditions for the time series to have a one-sided representation as in \eqref{wu} have not been established so far. In this work we consider the class of processes which admit the representation as in \eqref{wu} without providing further conditions which completely characterize this class. 

As targets of statistical inference we consider higher-order statistics contained in the following broad class of functions of generalized means as considered in Example 2.2 of \cite{kunsch1989jackknife}, Assumption C of \cite{buhlmann1997sieve} and Assumption (A2) of \cite{kreiss2011range}. Suppose we can observe univariate random variables $X_1, \hdots, X_n$ from some stationary process  $\bold{X} = \{X_t: t \in \mathbbm{Z}\}$. For functions $g: \mathbbm{R}^m \to \mathbbm{R}^d$ and $w: \mathbbm{R}^d \to \mathbbm{R}$ let 
\begin{align}\label{statistics}
T_n = w\left(\frac{1}{n-m+1} \sum_{t=1}^{n-m+1} g(X_t, \hdots, X_{t+m-1})\right),
\end{align}
where $m \in \mathbbm{N}$ and $d \in \mathbbm{N}$ are given numbers and the functions $g$ and $w$ fulfill some smoothness assumptions like in Assumption C of \cite{buhlmann1997sieve}. 
The so-defined class of statistics is quite rich and contains, e.\ g., sample means, sample autocovariances, sample autocorrelations, sample partial autocorrelations, and Yule-Walker estimators.

Under appropriate mixing or weak dependence conditions, central limit theorems for $T_n$ can be established for sufficiently smooth functions $g$ and $w$; cf., for example, \cite{kunsch1989jackknife}, \cite{buhlmann1997sieve}, \cite{kreiss2011bootstrap}, and \cite{jentsch2013}. However, in finite samples a normal approximation of the distribution of $T_n$ is often inaccurate and/or the limiting variance $\tau^2$ (say) is difficult to estimate or cannot be derived analytically. Therefore, and in line with previous literature, we suggest to employ a bootstrap procedure to approximate the unknown finite sample distribution of $T_n - \theta$, where $\theta$ is a centering constant or a parameter value under a null hypothesis of a statistical test, respectively. In particular, we base our bootstrap procedure on the Volterra representation \eqref{volterra} of the process (or a truncated version thereof) which can mimic higher-order moments of the process $\bold{X} =\{X_t: t \in \mathbbm{Z}\}$; see Example \ref{exam-autocorr} for a statistic which requires correctly mimicked fourth-order moments.

The remainder of the work is structured as follows. 
Section \ref{method} outlines the proposed Volterra-based procedure, Section \ref{theory} analyzes theoretical properties of the suggested procedure, Section \ref{estimation} explains how coefficients in the Volterra representation are estimated based on polynomial kernel regression, and Section \ref{MC} presents results of Monte-Carlo simulations highlighting the advantages of the suggested procedure over the autoregressive (AR) sieve bootstrap. Finally, Section \ref{discuss} concludes. 

\section{Volterra bootstrap}\label{method}
Before we describe our proposed methodology, let us consider a motivating example, which we will get back to 
in our numerical examples in Section \ref{MC}. 

\begin{example}[Sample autocorrelations at lag $1$]\label{exam-autocorr}
Consider the statistic $T_n$ from \eqref{statistics} for the special case of $m = d = 2$, $g(x, y) = (yx, x^2)^\top$, and $w(x, y) = x / y$. We obtain that
\begin{equation}\label{tn-autocorr1}
T_n = \frac{\sum_{t=1}^{n-1} X_{t+1} X_t}{\sum_{t=1}^{n-1} X_t^2}.
\end{equation}
Up to (empirical) centering, this statistic is for large sample size $n$ essentially equivalent to the sample autocorrelation $\hat{\rho}(1) = \widehat{\gamma}(1)/\widehat{\gamma}(0)$, where $\widehat{\gamma}(h) = n^{-1} \sum_{t=1}^{n - h} (X_t - \bar{X}_n) (X_{t+h} - \bar{X}_n)$ and $\bar{X}_n = n^{-1} \sum_{t=1}^n X_t$.

For convenience and due to practical relevance, we present here results pertaining to $\hat{\rho}(1)$, but they would apply in an analogous manner to $T_n$ from \eqref{tn-autocorr1}.
Namely, large sample properties of $\{\hat{\rho}(h)\}_{1 \leq h \leq k}$ for $k \in \mathbb{N}$ have been discussed by \cite{romano1996inference} under weak assumptions. In particular, the authors provided the following result.

\begin{propo}[Thm. 3.2 in \cite{romano1996inference}]\label{propo-romano}
Suppose $X_1, \hdots, X_n$ is a sample from a stationary mean zero process such that $\gamma(0) = \text{Var}(X_1) \in (0, \infty)$. Then, under appropriate moment and mixing conditions, the random vector $\sqrt n \left(\widehat{\rho}(1) - \rho(1), \hdots, \widehat{\rho}(k) - \rho(k)\right)^\top$ is asymptotically normal with mean vector zero. The asymptotic covariance $\tau_{i,j}$ of $\sqrt{n} \left(\widehat{\rho}(i) - \rho(i)\right)$ and $\sqrt{n} \left(\widehat{\rho}(j) - \rho(j)\right)$ is given by
\begin{eqnarray*}
\tau_{i,j} &\equiv &\lim_{n \to \infty} \{n \text{Cov}(\widehat{\rho}_n(i),\widehat{\rho}_n(j)) \} \\
& = &\gamma^{-2}(0) \left\{c_{i+1, j+1} - \rho(i) c_{1,j+1} - \rho(j) c_{1,i+1} + \rho(i) \rho(j) c_{1,1} \right\},
\end{eqnarray*}
where 
\begin{align*}
c_{i+1, j+1} &\equiv \lim_{n \to \infty} \{n \text{Cov}(\widehat{\gamma_n}(i),\widehat{\gamma_n}(j) ) \}\\
& \equiv \sum_{h = - \infty}^{\infty} \Big \{\gamma(h) \gamma(h+j-i) + \gamma(h+j) \gamma(h-i) + \kappa(h,i,j-i)   \Big \}\\
&= \sum_{h=-\infty}^{\infty} \text{Cov}(X_0 X_i, X_h X_{h+j})
\end{align*}
and $\kappa(h,i,j-i)$ denotes the fourth joint cumulant of the distribution of $(X_0, X_{i}, X_{h}, X_{j+h})^\top$.
\end{propo}
In the case that $\kappa(h,i,j-i)$ vanishes for all $(h, i, j)$, we arrive at Bartlett's formula (see, e.\ g., Theorem 7.2.1. in \cite{brockwell-davis1991}). However, this is only the case for restrictive special cases. For instance, Bartlett's formula is valid in the case that $\mathbf{X}$ is a Gaussian process or if $\mathbf{X}$ can be represented as a linear process of the form 
\begin{align}\label{linear}
X_t = \sum_{j=-\infty}^{\infty} b_j \varepsilon_{t-j}, \ b_0 = 1, t \in \Z,
\end{align}
where $\{\varepsilon_t: t \in \Z \}$ are  iid with zero mean and finite fourth moments, and the coefficients $\{b_j\}_{j \in \mathbb{Z}}$ are absolutely summable.
In general and for many processes of practical interest, however, fourth-order moments appear in the limiting (co-)variances $(\tau_{i,j})_{i,j}$.
\end{example}

\begin{rem}
The AR sieve bootstrap can only mimic first and second order moments of $\mathbf{X}$ correctly; see, e.\ g., Section 2.2.6. of \cite{kreiss2011bootstrap} and Section 3.1 of \cite{jentsch2013}.
\end{rem}

Our proposed methodology relies on a truncated version of the Volterra representation of the time series as in Equation \eqref{volterra}. For given order $p \in \mathbb{N}$ and degree $m \in \mathbb{N}$, it is given by 
\begin{align}\label{finite}
h^{(0)} + \sum_{u=0}^{m} h^{(1)} \varepsilon_{t-u} + \sum_{u=0}^{m} \sum_{v=0}^{m} h^{(2)} \varepsilon_{t-u} \varepsilon_{t-v} + \hdots + \sum_{u=0}^{m} \sum_{v=0}^{m} \hdots \sum_{w=0}^{m} h^{(p)} \varepsilon_{t-u} \varepsilon_{t-v} \hdots \varepsilon_{t-w}. 
\end{align}
Appropriate truncation is essential in the case of a finite sample size $n$.  A natural explanation provided by Volterra himself is that the memory [of a process] ''gradually fades out'', see \cite{volterra1959theory}. It is further formalized by \cite{boyd1985fading} and \cite{sandberg2002spl} among others. We provide details on an automated choice of $p$ and $m$ in Section \ref{MC}. 

Let $T_n$ be as in \eqref{statistics}, and suppose that for some appropriately increasing sequence of real numbers $\{c_n: n \in \N \}$ and a given real parameter $\theta$, the distribution $\mathcal{L}_n \equiv \mathcal{L} (c_n (T_n - \theta))$ has a nondegenerate limit. The Volterra bootstrap procedure to estimate the distribution $\mathcal{L}_n$ is then performed as follows.

\begin{algo}[Volterra bootstrap procedure]\label{algo-volterra-boot} $ $
\begin{enumerate}
\item Select an appropriate order $p << n$, and an appropriate degree $m << \infty$, and fit a $p$-th order $m$-th degree Volterra series to $X_1, \hdots, X_n$. The fitted process is denoted by $\hat{X}_t $ and is given as follows:
\begin{align}\label{step1}
\hat{X}_t = \sum_{u=0}^{m} \hat{h}^{(1)} \varepsilon_{t-u} + \sum_{u=0}^{m} \sum_{v=0}^{m} \hat{h}^{(2)} \varepsilon_{t-u} \varepsilon_{t-v} + \hdots +  \sum_{u=0}^{m} \sum_{v=0}^{m} \hdots  \sum_{w=0}^{m} \hat{h}^{(p)} \varepsilon_{t-u} \varepsilon_{t-v}\hdots \varepsilon_{t-w},
\end{align}
where the $\varepsilon_t$'s are iid, and $\hat{h}^{(\cdot)}$ is an estimated Volterra kernel of a corresponding order. For example, in Section \ref{MC} we consider $(\varepsilon_t)_t \overset{iid}{\sim} \mathcal{N}(0,1)$. 
\item Let $X_1^*, \hdots, X_n^*$ be constructed as follows:
\begin{align}\label{step2}
X^*_t = \sum_{u=0}^{m} \hat{h}^{(1)} \varepsilon^*_{t-u} + \sum_{u=0}^{m} \sum_{v=0}^{m} \hat{h}^{(2)} \varepsilon^*_{t-u} \varepsilon^*_{t-v} + \hdots +  \sum_{u=0}^{m} \sum_{v=0}^{m} \hdots  \sum_{w=0}^{m} \hat{h}^{(p)} \varepsilon^*_{t-u} \varepsilon^*_{t-v}\hdots \varepsilon^*_{t-w},
\end{align}
where $(\varepsilon_t^*)_t$ has the same (joint) distribution as $(\varepsilon_t)_t$.
\item Let $T^*_n = T_n(X_1^*, \hdots, X_n^*)$ be the statistic $T_n$ applied to the pseudo-time series $X_1^*, \hdots, X_n^*$, and denote by $\theta^*$ the analogue of $\theta$  associated with the bootstrap process 
$\mathbf{X}^*$. The Volterra bootstrap approximation of $\mathcal{L}_n$ is then given by $\mathcal{L}_n^* = \mathcal{L}^*(c_n(T_n^* - \theta^*))$, where $\mathcal{L}^*$ refers to the distribution of $(X_t^*)_{1 \leq t \leq n}$. In practice, a Monte Carlo-variant of $\mathcal{L}^*$ will be applied.
\end{enumerate}  
\end{algo}

In the next section we provide theoretical considerations regarding the consistency of the bootstrap  procedure defined by Algorithm \ref{algo-volterra-boot}. 

\section{Theoretical considerations}\label{theory}

In this section we present the key definitions and assumptions required in order to establish the consistency of the proposed Volterra bootstrap procedure. In this, the so-called cumulant matching approach as suggested by \cite{Kalouptsidis2005} plays an important role. 

\begin{defi}
We call a given bootstrap procedure, which generates pseudo observables $X_1^*, \hdots, X_{k(n)}^*$, consistent for $T_n$, if $d(\mathcal{L}_n, \mathcal{L}_n^*) \to 0$ in probability for $n \to \infty$, where $d(\cdot, \cdot)$ is any distance that metrizes weak convergence, e. g., the Prohorov distance. Here, $\{k(n)\}_{n \in \mathbb{N}}$ is an increasing sequence of integers which denotes bootstrap pseudo sample sizes.
\end{defi} 

The following assertion on the consistency of the bootstrap procedure is well-known in the bootstrap literature and has been discussed extensively by \cite{kreiss2011bootstrap}, among others. 

\begin{propo}[Conditions for bootstrap consistency] 
The consistency of a given bootstrap procedure for approximating $\mathcal{L}_n$ depends on the following two conditions.
\begin{itemize}
\item[(a)] The bootstrap procedure is such, that its resulting companion process (in the sense of \cite{kreiss2011bootstrap}) captures all distributional characteristics of $\mathbf{X}$ which are relevant for the limiting distribution of $c_n (T_n - \theta)$.
\item[(b)] The functions $g$ and $w$ are sufficiently smooth, such that distributional closeness of $\mathbf{X}$ and the companion process of the bootstrap procedure implies distributional closeness of $T_n$ and $T_n^*$. 
\end{itemize}
\end{propo}

The exact mathematical assumptions for smoothness of $g$ and $w$ can be found, e.\ g., under Assumption C of \cite{buhlmann1997sieve} and in $(A2)$ of \cite{kreiss2011range}, respectively. The following more explicit corollary is tailored to the setting of Example \ref{exam-autocorr}, or a similiar setting in which the limiting distribution of $c_n (T_n - \theta)$ is a normal distribution.

\begin{coro}\label{coro-cumulant-matching}
Assume that the functions $g$ and $w$ as well as the process $\mathbf{X}$ are such, that a central limit theorem holds for $c_n (T_n - \theta)$, where the limiting normal distribution is centered and its variance depends only on (joint) cumulants of finite order $\Xi \in \N$ of the distribution of $\mathbf{X}$. Then, a given bootstrap procedure for approximating $\mathcal{L}_n$ is consistent, if all (joint) cumulants up to order $\Xi$ of the distribution of $\mathbf{X}$ are correctly mimicked by the companion process of that bootstrap procedure.
\end{coro}

\cite{Kalouptsidis2005} provide the following results on the relationship between input cumulants and output cumulants of a Volterra system.

\begin{propo}[cf.\ Sections II and III of \cite{Kalouptsidis2005}]\label{propo-cumulant-matching}
Let $\Xi \in \N$ be a given integer and assume that the (random) input of a Volterra system is chosen to be stationary higher order white noise. Then there exist integers $m \in \mathbb{N}$ and $p \in \mathbb{N}$ as well as Volterra kernels $h^{(0)}, \hdots, h^{(p)}$, such that the (joint) cumulants up to order $\Xi $ of the finite Volterra series \eqref{finite} match given target values. 
\end{propo}

Proposition \ref{propo-cumulant-matching} guarantees that a cumulant matching up to a given order is possible by appropriately chosen Volterra kernels. In particular, this implies that the (joint) cumulants of our original process $(X_t)_t$ and the (joint) cumulants of the approximation $(\hat{X}_t)_t$ can be made identical by applying a suitable estimation procedure for Volterra kernels. Furthermore, since under \eqref{step2} in Step 2 of Algorithm \ref{algo-volterra-boot} we use the same (joint) distribution for 
$(\varepsilon^*_t)_t$ as for $(\varepsilon_t)_t$ in Step 1, we can deduce that the companion process corresponding to the bootstrap procedure defined by Algorithm \ref{algo-volterra-boot} can mimic the (joint) cumulants up to a required order $\Xi$ of the original process $(X_t)_t$. This argumentation implies the conceptual validity of the proposed Volterra bootstrap approach under the assumptions of Corollary \ref{coro-cumulant-matching}. 

It remains to describe an appropriate estimation and model selection procedure for $m$, $p$, and $h^{(0)}, \hdots, h^{(p)}$.  In the following section we employ a technique which is based on the theory of 
reproducing kernel Hilbert space (RKHS) and polynomial kernel regression. The reason for this choice is that this estimation method scales linearly with the input dimensionality and is independent of the degree of nonlinearity. This avoids stability issues (cf., e.\ g., \cite{franz2005implicit, franzschoelkopf2006}) of direct cumulant matching approaches, especially for larger values of $m$ and $p$. 


\section{Estimation approach}\label{estimation}

Several methods to estimate Volterra kernels exist in the literature. Among others, there are the cross-correlation method by \cite{lee1965measurement} and its extensions such as, e.\ g., in \cite{orcioni2018identification}, the exact orthogonal method as in \cite{korenberg1996identification}, the neural network-based method as in \cite{wray1994calculation} and the polynomial kernel regression method as in \cite{franz2005implicit}. The cross-correlation method is considered to be a traditional method to estimate the Volterra representation and is widely applied. However, as outlined by \cite{franz2005implicit}, it suffers from several shortcomings: (1) It requires large sample sizes before sufficient convergence is reached. (2) Generally (and initially) it is developed under the assumption of Gaussian iid inputs. (3) The number of coefficients to be estimated for the finite-sample Volterra expansion is $(p+m-1)!/(p!(m-1)!)$, which can be computationally prohibitive already in moderately scaled models. (4) Estimation is performed under the noise-free data assumption which is unrealistic as real data is likely to be noise-contaminated, see Section 2 in \cite{franz2005implicit}. 

For these reasons we adopt the estimation method suggested by \cite{franz2005implicit}, which overcomes the disadvantages of the cross-correlation method as listed above and can provide estimates of the Volterra kernels in a much more (computationally) efficient way. The key idea of \cite{franz2005implicit} consists in reformulating the Volterra series as a polynomial kernel regression in a RKHS. In the remainder of this section we provide a summary on this estimation method. Further details can be found in \cite{franz2004-IEEE, franz2005implicit, franzschoelkopf2006} and references therein. We use bold letters to denote vectors and matrices, respectively. 

It is convenient to explain polynomial kernel regression in RKHS by starting with the linear regression. Assume that the process $\bf X$ is approximated as a function of $\boldsymbol{\varepsilon}$, meaning that the following representation holds:
\begin{align}\label{linear2}
\hat{X}_t = f(\boldsymbol{\varepsilon}_t) = \sum_{j=0}^M \gamma_j \varphi_j (\boldsymbol{\varepsilon}_t),  \ 1 \leq t \leq n,
\end{align}
where $\boldsymbol{\varepsilon}_t = (\varepsilon_t, \hdots, \varepsilon_{t-m+1}) \in \R^m$, $\gamma_j \in \R$, $\varphi_j: \R^{m} \to \R$ and $\varphi_0 (\boldsymbol{\varepsilon}_t)= 1$, and where the $\varphi_j$'s contain all monomials of the elements of the vector $\boldsymbol{\varepsilon}_t$ up to order $j$ for the $j$-th order Volterra series. The coefficients $\{\gamma_j\}_{0 \leq j \leq M}$ are found by minimizing the mean squared error (MSE) as follows:
\begin{align}\label{linearMSE}
\widehat{\boldsymbol{\gamma} } = \arg \min_{\boldsymbol{\gamma}} n^{-1} \sum_{t=1}^n (\hat{X}_t  - X_t)^2,
\end{align}
where $\widehat{\boldsymbol{\gamma}} = (\widehat{\gamma}_0, \hdots, \widehat{\gamma}_M)$. Since the number of coefficients to be estimated for the $p$-th order $m$-th degree Volterra expansion is $(p+m-1)!/(p!(m-1)!)$, the linear regression approach might no longer be computationally efficient, whereas if one employs the polynomial kernel regression framework instead of the $M$ functions $\varphi_1, \hdots, \varphi_M$, the computations can be carried out much faster. In what follows we show how Volterra series can be rewritten as a linear operator in a RKHS. 

First, we rewrite \eqref{finite} as a sum of Volterra operators as follows: 
\begin{align}\label{rewrite}
\hat{X}_t = f(\boldsymbol{\varepsilon}_t) =  \sum_{i=0}^p H_i (\boldsymbol{\varepsilon}_t), \ 1 \leq t \leq n,
\end{align}
where $H_i (\boldsymbol{\varepsilon}_t) = \sum_{j_1 = 1}^m \hdots  \sum_{j_i = 1}^m h^{(i)}_{j_1, \hdots, j_i}\varepsilon_{j_1} \hdots \varepsilon_{j_i}$ is the $i$-th order Volterra operator. Further we define the following maps:
\[\phi_0(\boldsymbol{\varepsilon}_t) = 1 \ \text{and} \ \phi_i(\boldsymbol{\varepsilon}_t ) = (\varepsilon_t^i, \varepsilon_{t}^{i-1} \varepsilon_{t-1}, \hdots, \varepsilon_t \varepsilon_{t-1}^{i-1}, \varepsilon_{t-1}^i, \hdots, \varepsilon^i_{t-m+1}), \ 0 \leq i \leq p,\]
such that $\phi_i$ maps the input $\boldsymbol{\varepsilon}_t \in \R^m$ into a vector $\phi_i(\boldsymbol{\varepsilon}_t) \in \mathbbm{R}^{m^i}$. By stacking the coefficients of the $i$-th order Volterra operator into a single vector $\boldsymbol{\eta}_i = (h^{(i)}_{1,1,\hdots,1}, h^{(i)}_{1,2,\hdots,1}, \hdots) \in \mathbbm{R}^{m^i}$ we can rewrite it as a scalar product as follows: 
\[H_i (\boldsymbol{\varepsilon}_t) = \boldsymbol{\eta}_i^\top \phi_i(\boldsymbol{\varepsilon}_t),  \ 0 \leq i \leq p.\]
Finally, we stack the maps $\phi_i$ with positive weights $a_i \in \R_{>0}$ into a single map $\phi^{(p)}(\boldsymbol{\varepsilon}_t) = (a_0 \phi_0 (\boldsymbol{\varepsilon}_t), a_1 \phi_1 (\boldsymbol{\varepsilon}_t), \hdots, a_p \phi_p (\boldsymbol{\varepsilon}_t))^\top$, where $\phi^{(p)}(\boldsymbol{\varepsilon}_t): \ \R^m \to \R \times \R^m \times \R^{m^2} \times \hdots \times \R^{m^p} = \R^M$ and $M = (1-m^{p+1})/(1-m)$. It follows that Equation \eqref{rewrite} can be rewritten as a scalar product as follows
\begin{align}\label{rkhsReg}
\hat{X}_t =  f(\boldsymbol{\varepsilon}_t) = \sum_{i=0}^p H_i (\boldsymbol{\varepsilon}_t) = (\boldsymbol{\eta}^{(p)})^\top \phi^{(p)}(\boldsymbol{\varepsilon}_t), \ 1 \leq t \leq n,
\end{align}
where $\boldsymbol{\eta}^{(p)} \in \R^M$. Similar to Equation \eqref{linearMSE} the optimal solution can be expressed as follows:
\begin{align}\label{volterraMSE}
\widehat{\boldsymbol{\eta}}^{(p)} = \arg \min_{\boldsymbol{\eta}^{(p)}} n^{-1} \sum_{t=1}^n (f(\boldsymbol{\varepsilon}_t)  - X_t)^2 + \lambda (\boldsymbol{\eta}^{(p)})^{\top} \boldsymbol{\eta}^{(p)},
\end{align}
where $\lambda$ is additionally introduced as a regularizing penalty, which accounts for the noise in the real data and can be determined in practice, e.\ g., via cross-validation.  This solution is not yet based on kernels and is computationally no more efficient than the solution to Equation \eqref{linear2}. However, by reformulating \eqref{finite} as in \eqref{rkhsReg} one can employ the fact that the space of functions $\phi_i (\boldsymbol{\varepsilon}_t)$, $i = 0, \hdots, p$, has the structure of a RKHS, see \cite{scholkopf2002learning}. Namely, it can be shown that
\[\phi_i (\boldsymbol{\varepsilon}_t)^\top \phi_i (\boldsymbol{\varepsilon}_{t'}) = (\boldsymbol{\varepsilon}_t^\top \boldsymbol{\varepsilon}_{t'})^i  \equiv k_i(\boldsymbol{\varepsilon}_t,  \boldsymbol{\varepsilon}_{t'}), \ 1 \leq t, \ t' \leq n,\]
where $k_i(\boldsymbol{\varepsilon}_t, \boldsymbol{\varepsilon}_{t'})$ is the $i$-th degree homogeneous polynomial kernel. Consequently, one can also write the scalar product of the maps $\phi^{(p)}(\boldsymbol{\varepsilon}_t)$ as follows:
\[\phi^{(p)}(\boldsymbol{\varepsilon}_t)^\top \phi^{(p)}(\boldsymbol{\varepsilon}_{t'}) = \sum_{i=0}^p a^2_i  (\boldsymbol{\varepsilon}_t^\top \boldsymbol{\varepsilon}_{t'})^i  \equiv k^{(p)}(\boldsymbol{\varepsilon}_t,  \boldsymbol{\varepsilon}_{t'}), \ 1 \leq t, \ t' \leq n.\]
Due to the RKHS structure of the space of the functions  $\phi_i (\boldsymbol{\varepsilon}_t)$, $i = 0, \hdots, p$, it follows from the representer theorem that the optimal solution to Equation \eqref{rkhsReg} can be expressed in terms of kernels as follows: 
\begin{align}\label{rkhsKernels}
\hat{X}_t =  f(\boldsymbol{\varepsilon}_t) =  \sum_{i=0}^p H_i (\boldsymbol{\varepsilon}_t) = \bold{X}^\top \  (\bold{K}_p + \lambda \bold{I}_n)^{-1} \ \bold{k}^{(p)} (\boldsymbol{\varepsilon}_t), \ 1 \leq t \leq n,
\end{align}
where $\bold{X} = (X_1, \hdots, X_n)$ denotes a $n \times 1$ vector, $\bold{K}_p$ is the (positive definite) $n \times n$ Gram matrix with entries $k^{(p)}(\boldsymbol{\varepsilon}_t,  \boldsymbol{\varepsilon}_{t'}), \ 1 \leq t, \ t' \leq n$, and $\bold{k}^{(p)} (\boldsymbol{\varepsilon}_t) \in \mathbb{R}^{n \times 1}$ denotes the $t$-th column of $\bold{K}_p$, $1 \leq t \leq n$.

To recover the coefficients of each Volterra kernel individually, note that the coefficient vector $\boldsymbol{\eta}_i = (h^{(i)}_{1,1,\hdots,1}, h^{(i)}_{1,2,\hdots,1}, \hdots)^\top$ of the $i$-th order Volterra operator can equivalently written as follows:
\[\boldsymbol{\eta}_i = a_i \boldsymbol{\Phi}_i^\top (\bold{K}_p + \lambda \bold{I}_n)^{-1} \mathbf{X}, \ 1 \leq i \leq p,\]
where $\boldsymbol{\Phi}_i = (\phi_i (\boldsymbol{\varepsilon}_1), \hdots, \phi_i (\boldsymbol{\varepsilon}_n))^\top$ is 
the matrix containing all monomials corresponding to the $i$-th order Volterra operator. 

The choice of an appropriate penalty $\lambda$ as well as the choice of an order $p$ and a degree $m$ of the finite (truncated) Volterra representation can be performed either by minimizing the in-sample MSE of a corresponding fit for each possible $\lambda$, $p$ and $m$, or by cross-validation in the frequency-domain as suggested by \cite{hurvich1990frequency}. The latter approach is computationally much more intensive, however, less biased, whereas the former is faster, but often leads to overfitting due to the fact that the crucial assumption of cross-validation on independence of test and training sets is not valid for time series.

\section{Simulation studies}\label{MC}

In this section we summarize results of our simulation studies. We consider a linear process and several nonlinear processes to illustrate the performance of the suggested Volterra bootstrap procedure for the case of testing for autocorrelation at lag 1 based on the estimator $\widehat{\rho}(1)$; see Example \ref{exam-autocorr}. To highlight the usefulness of the suggested procedure  we also perform the AR sieve bootstrap for the processes under consideration. 

In particular, we consider two-sided test problems of the form
\[
H_0: \rho(1) = c_0 \text{~~versus~~} H_1: \rho(1) \neq c_0
\]
for a given value $c_0 \in [-1, 1]$. The accuracy of the approximation of the null distribution of $\widehat{\rho}(1)$ by means of the Volterra bootstrap is assessed by reporting empirical type I error rates (i.\ e., relative rejection frequencies) of the hypothesis test which is given by the following scheme.

\begin{algo}\label{algo-test-lag1} $ $
\begin{enumerate}
\item Fix the significance level $\alpha$ of the test, and fix a number $B$ of bootstrap repetitions. 
\item Let a Studentized version of the absolute difference between $\widehat{\rho}(1)$ and $c_0$ be given by
\[
D_n = \sqrt{n} \left|\frac{\widehat{\rho}(1) - c_0}{\sqrt{\widehat{\text{Var}}(\widehat{\rho}(1))}}\right|,
\]
and let $D_n^{*, b}$ denote the analogue of $D_n$ based on the Volterra bootstrap process $\mathbf{X}^*$ according to Section \ref{method} in the $b$-th bootstrap repetition.
\item Let a bootstrap $p$-value for testing $H_0$ versus $H_1$ be given by
\[
p_{\text{boot}} = \frac{|\{b: D_n^{*, b} > D_n\}| + 1}{B + 1}.
\]
\item Reject $H_0$ in favor of $H_1$ iff $p_{\text{boot}} < \alpha$.
\end{enumerate}
\end{algo}

In our simulations, we have set $c_0 = \rho(1)$, meaning that the null hypothesis $H_0$ is true, and we have set $\alpha = 5\%$.
In analogy to \cite{jentsch2013}, the true autocorrelation has been approximated by means of $20{,}000$ Monte-Carlo simulations for each of the processes under consideration. The variance of $\widehat{\rho}(1)$ has been estimated using the formulas in Proposition \ref{propo-romano} based on 10 lead and 10 lags of the simulated process. For the AR sieve bootstrap we used the Akaike information criterion to fit the model and $p_{max} = 20$. 

We consider the following processes: 
\begin{enumerate}
\item [P1]
\textbf{AR}, $X_t = 0.75 X_{t-1} + \varepsilon_t, \ \varepsilon_t \overset{iid}{\sim} \mathcal{N}(0,1)$.
\item [P2]
\textbf{GARCH}, $X_t = \sigma_t \varepsilon_t$, $\sigma^2_t = 1 + 0.2 \sigma^2_{t-1} + 0.65 \varepsilon^2_{t-1}$, $\varepsilon_t \overset{iid}{\sim} \mathcal{N}(0,1)$.
\item [P3]
\textbf{Bilinear}, $X_t = 0.6 X_{t-1} + \varepsilon_t + 0.75 X_{t-1}\varepsilon_{t-1}$,  \ $\varepsilon_t \overset{iid}{\sim} \text{UNI}(-\sqrt 3, \sqrt 3)$.
\item [P4]
\textbf{EXPAR}, $X_t = (0.45 + 0.48 \exp(-0.96 X^2_{t-1}))X_{t-1} + \varepsilon_t$, $\varepsilon_t \overset{iid}{\sim} \text{UNI}(-\sqrt 3, \sqrt 3)$.
\end{enumerate}

Each of the processes is generated under stationarity assumptions as stated, e. g., in \cite{wu2011asymptotic}. When simulating the corresponding time series as well as in the bootstrap procedure we have skipped the first $N$ values until the process achieves stationarity, where $N = 100$ is typically sufficient. In expressions as, for instance, the right-hand side of \eqref{volterraMSE}, a corresponding shift of the time index has to be considered.

For each model P1-P4 we consider time series of length $n=100$ and 200 simulation runs with 250 bootstrap repetitions within each simulation run to assess the empirical type I error rate of the proposed bootstrap test. For all processes, we used $(\varepsilon_t)_t \overset{iid}{\sim} \mathcal{N}(0,1)$ as well as
$(\varepsilon^*_t)_t \overset{iid}{\sim} \mathcal{N}(0,1)$ in estimation and bootstrapping based on the Volterra representation. The order $p$ and the degree $m$ of the Volterra representation have been chosen either based on minimizing the in-sample MSE or based on the procedure as in \cite{hurvich1990frequency}. For regularized estimation as explained in Section \ref{estimation} we used the following penalties: $\lambda \in \{10e-7, 10e-6, 10e-5\}$. The maximum degree in the Volterra representation is set equal to $30$, and the maximum order is set equal to $p_{max} = 10$. We report our simulation results in Table \ref{MCresults}.

Our simulation results for P1 for the AR sieve procedure are as expected from the theoretical point of view, i.\ e., given that the innovations are Gaussian, the AR sieve boostrap performs well. However, for the processes P3 - P4 this is not the case anymore as either the innovations are not Gaussian, or the process under consideration is no longer representable as in Equation \eqref{linear}. Interestingly, also for P2 when the process is nonlinear, but the innovations are Gaussian, the test based on AR sieve seems to violate the type I error rate in finite samples. On the other hand, the Volterra bootstrap is able to keep the type I error rate approximately at the pre-specified significance level for nonlinear processes due to its ability to replicate the higher-order structure of the underlying process. However, for the linear processes unless the order of the Volterra representation is restricted to one, the type I error rate is slightly larger than $\alpha$, because the values of $p$ and $m$ chosen both by cross-validation and by in-sample MSE minimization result in overfitting. It is therefore necessary before deciding on the appropriate bootstrapping scheme to test the time series under consideration for nonlinearity. 

\begin{center}
\begin{table}[ht!]
\caption{Type I errors (significance level $5\%$) for AR sieve and Volterra bootstrap procedures, $n = 100$, the number of Monte-Carlo repetitions is 200 and the number of bootstrap repetitions is 250. The symbol $*$ indicates that the order of Volterra representation was restricted to one.}\label{MCresults}
\centering
\begin{tabular}{ c cc cc} 

\hline
& & & & \\
Model & AR sieve  & Average $p$ & Volterra & Average ($p, \ m$) \\[0.25cm]
\hline
& & & & \\
P1 & 0.048 & 1.0 & 0.042 & ($1.0^{*}$, 30.0) \\[0.25cm]
P2 & 0.084 & 1.4 & 0.057 & (2.8, 29.2)\\[0.25cm]
P3 & 0.068 & 1.2 & 0.042 & (2.1, 22.4)\\[0.25cm]
P4 & 0.059 & 1.0  & 0.052 & (2.4, 28.2)\\
& & & & \\
\hline
\end{tabular}
\end{table}
\end{center}
%

\section{Discussion}\label{discuss}

In the present work we focus on the bootstrap procedure based on the Volterra series representation. In particular, we estimate and mimic the original process based on iid random variables. An alternative procedure can be constructed based on the lags of the original process similarly as in the AR sieve bootstrap method. For example, consider the following representation:
\[X_t = f(X_{t-1}, \varepsilon_t),\]
where $f$ is some measurable function such that $X_t$ is well-defined. This type of approach has been indicated by \cite{closed-loop-Nature}. To our knowledge the necessary and sufficient conditions for the existence of such a representation have not been established so far. In \cite{wiener1958nonlinear} the problem of finding a so-called (infinite) nonlinear moving average representation is dealt with in Lecture 11 ("Coding"), whereas an (infinite) autoregressive representation is addressed in Lecture 12 ("Decoding"). A somewhat more detailed discussion of these ideas is available in \cite{kallianpur1981some} and \cite{masani1966wiener}. However, neither \cite{wiener1958nonlinear} nor \cite{masani1966wiener} established nonlinear AR-type filtering theory in full detail. A so called  ''coefficient matching'' approach with the goal of generalizing linear AR results is attempted in \cite{hunt1995nonlinear}. Furthermore, a nonlinear autoregressive representation for the case when the process under consideration is a Markov chain is worked out in \cite{rosenblatt2012markov}, see also \cite{tong1990non}. 

Another direction for formulating bootstrap procedures for nonlinear processes might be to consider its frequency domain representation employing higher-order spectra as in \cite{brillinger1970identification}, \cite{brillinger1994some}, \cite{shiryaev1960some}, \cite{shiryaev1963conditions}, and 
\cite{priestley1988non}. 

We reserve these ideas as well as practical applications of the suggested bootstrap procedure for future research. 

%
%
%


\end{document}